\documentclass[a4paper,preprintnumbers,amsmath,amssymb,nofootinbib]{revtex4}

\usepackage{graphicx}
\usepackage{dcolumn}
\usepackage{bm}
\usepackage{epstopdf}

\newcount\driver
\newcount\bozza

\font\ottorm=cmr8 scaled\magstep1 
scaled\magstep1                  
scaled\magstep1 \font\msytw=msbm10 scaled\magstep1
 
scaled\magstep1                 \font\indbf=cmbx10 scaled\magstep2

scaled \magstep2

\newcount\driver
\newcount\bozza

\font\ottorm=cmr8 scaled\magstep1 
scaled\magstep1                  
scaled\magstep1 \font\msytw=msbm10 scaled\magstep1
 
scaled\magstep1                 \font\indbf=cmbx10 scaled\magstep2

scaled \magstep2

{\count255=\time\divide\count255 by 60
\xdef\hourmin{\number\count255}
        \multiply\count255 by-60\advance\count255 by\time
   \xdef\hourmin{\hourmin:\ifnum\count255<10 0\fi\the\count255}}

\let\a=\alpha \let\b=\beta    \let\g=\gamma     \let\d=\delta     \let\e=\varepsilon
       \let\th=\vartheta \let\k=\kappa     \let\l=\lambda
\let\m=\mu    \let\n=\nu                      \let\r=\rho
\let\s=\sigma \let\t=\tau            
\let\ps=\psi

\def\PP{{\cal P}}\def\EE{{\cal E}}\def\VV{{\cal V}}
\def\HH{{\cal H}}\def\WW{{\cal W}}
\def\TT{{\cal T}}\def\BB{{\cal B}}\def\ZZ{{\cal Z}}
\def\RR{{\cal R}}\def\LL{{\cal L}}

\def\xx{{\bf x}}
   
\def\nn{{\bf n}}
\def\uu{{\bf u}}
 \def\bP{{\bf P}}
\def\tt{{\bf t}}\def\bT{{\bf T}}

       \def\oo{{\underline \omega}}
\def\ee{{\underline \varepsilon}}

\def\uu{\bf u}

\def\RRR{\hbox{\msytw R}}

        \def\EE{\hbox{\msytw E}}

\let\bs=\backslash
\let\==\equiv

\let\io=\infty
\let\0=\noindent

\def\*{{\hfill\break\null\hfill\break}}

\def\tilde#1{{\widetilde #1}}

\def\tende#1{\,\vtop{\ialign{##\crcr\rightarrowfill\crcr
             \noalign{\kern-1pt\nointerlineskip}
             \hskip3.pt${\scriptstyle #1}$\hskip3.pt\crcr}}\,}
\def\otto{\,{\kern-1.truept\leftarrow\kern-5.truept\to\kern-1.truept}\,}

\def\wh#1{\widehat{#1}}
\def\hat#1{\wh{#1}}
\def\sqt[#1]#2{\root #1\of {#2}}

\def\bp{{\bar \ps}}

\def\PP{{\cal P}}\def\EE{{\cal E}}\def\VV{{\cal V}}
\def\HH{{\cal H}}\def\WW{{\cal W}}
\def\TT{{\cal T}}\def\BB{{\cal B}}\def\ZZ{{\cal Z}}
\def\RR{{\cal R}}\def\LL{{\cal L}}

\def\T#1{{#1_{\kern-3pt\lower7pt\hbox{$\widetilde{}$}}\kern3pt}}
\def\VVV#1{{\underline #1}_{\kern-3pt
\lower7pt\hbox{$\widetilde{}$}}\kern3pt\,}
\def\W#1{#1_{\kern-3pt\lower7.5pt\hbox{$\widetilde{}$}}\kern2pt\,}

\def\indica{\leaders \hbox to 0.5cm{\hss.\hss}\hfill}
\def\guida{\leaders\hbox to 1em{\hss.\hss}\hfill}
\mathchardef\oo= "0521

\def\xx{{\bf x}}
\def\nn{{\bf n}}
\def\uu{{\bf u}}
 \def\bP{{\bf P}}
\def\tt{{\bf t}} 
\def\oo{{\underline \omega}}

\def\qed{\raise1pt\hbox{\vrule height5pt width5pt depth0pt}}
 
  \def\bp{{\bar p}} 

\def\indic{\hbox{\raise-2pt \hbox{\indbf 1}}}

\def\RRR{\hbox{\msytw R}}

\def\ins#1#2#3{\vbox to0pt{\kern-#2 \hbox{\kern#1 #3}\vss}\nointerlineskip}

\newdimen\xshift \newdimen\xwidth \newdimen\yshift

\def\insertplot#1#2#3#4#5#6{%
\xwidth=#1pt \xshift=\hsize \advance\xshift by-\xwidth \divide\xshift by 2%
\begin{figure}[ht]
\vspace{#2pt} \hspace{\xshift}
\begin{minipage}{#1pt}
#3 \ifnum\driver=1 \griglia=#6
\ifnum\griglia=1 \openout13=griglia.ps \write13{gsave .2
setlinewidth} \write13{0 10 #1 {dup 0 moveto #2 lineto } for}
\write13{0 10 #2 {dup 0 exch moveto #1 exch lineto } for}
\write13{stroke} \write13{.5 setlinewidth} \write13{0 50 #1 {dup 0
moveto #2 lineto } for} \write13{0 50 #2 {dup 0 exch moveto #1
exch lineto } for} \write13{stroke grestore} \closeout13
\includegraphics{griglia.ps} \fi
\includegraphics{#4.ps}\fi%
\ifnum\driver=2 \fi
\end{minipage}
\caption{#5}
\end{figure}
}

\newdimen\shift \shift=-1.5truecm
\def\lb#1{%
\ifnum\bozza=1
\label{#1}\rlap{\hbox{\hskip\shift$\scriptstyle#1$}}
\else\label{#1} \fi}

\def\be{\begin{equation}}
\def\ee{\end{equation}}
\def\bea{\begin{eqnarray}}\def\eea{\end{eqnarray}}
\def\bean{\begin{eqnarray*}}\def\eean{\end{eqnarray*}}
\def\bfr{\begin{flushright}}\def\efr{\end{flushright}}
\def\bc{\begin{center}}\def\ec{\end{center}}
\def\bal{\begin{align}}\def\eal{\end{align}}
\def\ba#1{\begin{array}{#1}} \def\ea{\end{array}}
\def\bd{\begin{description}}\def\ed{\end{description}}
\def\bv{\begin{verbatim}}\def\ev{\end{verbatim}}
\def\nn{\nonumber}
\def\Halmos{\hfill\vrule height10pt width4pt depth2pt \par\hbox to \hsize{}}
\def\pref#1{(\ref{#1})}

\def\ins#1#2#3{\vbox to0pt{\kern-#2 \hbox{\kern#1 #3}\vss}\nointerlineskip}

\newdimen\xshift \newdimen\xwidth \newdimen\yshift
\newcount\griglia

\def\insertplot#1#2#3#4#5#6{%
\xwidth=#1pt \xshift=\hsize \advance\xshift by-\xwidth \divide\xshift by 2%
\begin{figure}[ht]
\vspace{#2pt} \hspace{\xshift}
\begin{minipage}{#1pt}
#3 \ifnum\driver=1 \griglia=#6
\ifnum\griglia=1 \openout13=griglia.ps \write13{gsave .2
setlinewidth} \write13{0 10 #1 {dup 0 moveto #2 lineto } for}
\write13{0 10 #2 {dup 0 exch moveto #1 exch lineto } for}
\write13{stroke} \write13{.5 setlinewidth} \write13{0 50 #1 {dup 0
moveto #2 lineto } for} \write13{0 50 #2 {dup 0 exch moveto #1
exch lineto } for} \write13{stroke grestore} \closeout13
\includegraphics{griglia.ps} \fi
\includegraphics{#4.ps}\fi%
\ifnum\driver=2 \fi
\end{minipage}
\caption{#5}
\end{figure}
}

\newdimen\shift \shift=-1.5truecm
\def\lb#1{%
\label{#1}\rlap{\hbox{\hskip\shift$\scriptstyle#1$}}
\else\label{#1} \fi}

\def\be{\begin{equation}}
\def\ee{\end{equation}}
\def\bea{\begin{eqnarray}}\def\eea{\end{eqnarray}}
\def\bean{\begin{eqnarray*}}\def\eean{\end{eqnarray*}}
\def\bfr{\begin{flushright}}\def\efr{\end{flushright}}
\def\bc{\begin{center}}\def\ec{\end{center}}
\def\bal{\begin{align}}\def\eal{\end{align}}
\def\ba#1{\begin{array}{#1}} \def\ea{\end{array}}
\def\bd{\begin{description}}\def\ed{\end{description}}
\def\bv{\begin{verbatim}}\def\ev{\end{verbatim}}
\def\nn{\nonumber}
\def\Halmos{\hfill\vrule height10pt width4pt depth2pt \par\hbox to \hsize{}}
\def\pref#1{(\ref{#1})}

\def\ins#1#2#3{\vbox to0pt{\kern-#2 \hbox{\kern#1 #3}\vss}\nointerlineskip}

\newdimen\xshift \newdimen\xwidth \newdimen\yshift
\newcount\griglia

\def\insertplot#1#2#3#4#5#6{%
\xwidth=#1pt \xshift=\hsize \advance\xshift by-\xwidth \divide\xshift by 2%
\begin{figure}[ht]
\vspace{#2pt} \hspace{\xshift}
\begin{minipage}{#1pt}
#3 \ifnum\driver=1 \griglia=#6
\ifnum\griglia=1 \openout13=griglia.ps \write13{gsave .2
setlinewidth} \write13{0 10 #1 {dup 0 moveto #2 lineto } for}
\write13{0 10 #2 {dup 0 exch moveto #1 exch lineto } for}
\write13{stroke} \write13{.5 setlinewidth} \write13{0 50 #1 {dup 0
moveto #2 lineto } for} \write13{0 50 #2 {dup 0 exch moveto #1
exch lineto } for} \write13{stroke grestore} \closeout13
\includegraphics{griglia.ps} \fi
\includegraphics{#4.ps}\fi%
\ifnum\driver=2 \fi
\end{minipage}
\caption{#5}
\end{figure}
}

\usepackage{amsmath}
\usepackage{amsfonts}
\usepackage{amssymb}
\usepackage{epstopdf}

\font\msytw=msbm9 scaled\magstep1 
scaled\magstep1

\let\a=\alpha \let\b=\beta  \let\g=\gamma  \let\d=\delta
\let\e=\varepsilon
     \let\th=\theta \let\k=\kappa \let\l=\lambda
\let\m=\mu    \let\n=\nu             \let\r=\rho
\let\s=\sigma \let\t=\tau    
\let\ps=\Psi

\def\EE{{\cal E}} \def\VV{{\cal V}}
 \def\WW{{\cal W}}
\def\TT{{\cal T}} 
\def\RR{{\cal R}}\def\LL{{\cal L}}

 \def\xx{{\bf x}}  

\def\PP{{\bf P}}

\def\nn{\nonumber}

\def\RRR{\hbox{\msytw R}}

\def\\{\hfill\break}
\def\={:=}
\let\io=\infty
\let\0=\noindent

\def\tende#1{\,\vtop{\ialign{##\crcr\rightarrowfill\crcr\noalign{\kern-1pt
    \nointerlineskip} \hskip3.pt${\scriptstyle #1}$\hskip3.pt\crcr}}\,}
\def\otto{\,{\kern-1.truept\leftarrow\kern-5.truept\to\kern-1.truept}\,}

\def\wh{\widehat}
\def\to{\rightarrow}

\def\qed{\hfill\raise1pt\hbox{\vrule height5pt width5pt depth0pt}}

\def\be{\begin{equation}}
\def\ee{\end{equation}}
\def\bp{\begin{pmatrix}}
\def\ep{\end{pmatrix}}
\def\bea{\begin{eqnarray}}
\def\eea{\end{eqnarray}}
\def\nn{\nonumber}
\def\pref#1{(\ref{#1})}

\def\lb{\label}

\begin{document}

\title{Anomaly cancellation condition in an effective non-perturbative electroweak theory}
\author{Vieri Mastropietro}
\address{Dipartimento di Matematica, Universit\`a di Milano \\
  \small{Via Saldini, 50, I-20133 Milano, ITALY }}
\email{vieri.mastropietro@unimi.it}

\begin{abstract} 
We establish the non-perturbative validity of the gauge anomaly cancellation condition in an effective electroweak theory of massless fermions with finite momentum cut-off
and Fermi interaction.
The requirement that the current is conserved up to terms smaller than the energy divided by the cut-off scale, which is the natural condition as gauge invariance is only emerging,
produces the same constraint on charges as in the Standard Model.
The result holds at a non-perturbative level as
the functional integrals are expressed by convergent power series expansions and are analytic in a finite domain. 
\end{abstract}

\maketitle

\section{Introduction}

In a chiral theory classical gauge invariance can be broken 
at the quantum level by {\it anomalies}, a fact producing
a lack of renormalizability and of internal consistency. Therefore in the Standard Model 
the anomalies need to cancel out, and this produces
an algebraic condition on the hypercharges (and therefore on the charges), see e.g. \cite{R},  \cite{R1}
which gives a partial explanation to charge quantization without reference to grand unification.
The values of hypercharges, that at a classical level can take any value, are constrained by a purely quantum effect.

In order to see how this condition arises 
the 
gauge fields can be decomposed 
in classical background and quantum part, see e.g. \cite{T}. Neglecting the effect of the 
quantum gauge fields (but keeping quantum fermions)
the conservation of current, following classically by Noether theorem, is broken by 
terms quadratic in the fields proportional to the three current correlations.
They are expressed by sum of a small number of terms ( "triangle graphs")
and the 
condition on the charges comes from the request that such sum is vanishing.
When the effect of quantum gauge fields is taken into account, the anomaly is instead sum of infinitely many graphs. Radiative corrections could 
produce extra conditions but the 
Adler-Bardeen theorem \cite{AB} is invoked to say that this is not the case. 
Such a property, which says that the anomaly is not renormalized by interactions, is a perturbative statement relying on cancellations 
valid assuming exact symmetries and removed cut-off. Non--perturbative versions
of such property in a functional integral framework using the Jacobian method \cite{F} are indeed valid only at one loop, see e.g. \cite{K} or \cite{AB1}. 

Natural questions are if the cancellation condition is valid at a non-perturbative level and if it still holds
in an effective description when some symmetry is only approximate. 
The two questions are related; we can consider a
theory with finite cut-off which can be possibly studied non-perturbatively, but some symmetry
is necessarily broken. 
Such problems have been extensively analyzed over the years.
Lattice gauge theory is a natural framework for a non-perturbative construction
but the program of getting an anomaly free formulation of electroweak theory is still non complete, see for instance  
\cite{A0}-\cite{A3}.
In \cite{Pr},\cite{Pr1} the anomaly in a theory with finite momentum cut-off have been considered at a perturbative level.
The role of symmetry breaking terms in the anomaly cancellation was considered 
in \cite{P0}, finding that at one loop they do not break the cancellation condition.

In this paper we follow a different point of view. 
We keep a {\it finite} momentum ultraviolet cut-off and we
still decompose the gauge field 
in a classical background and a quantum field;
this second field  is integrated out
to produce an effective fermionic interaction. We consider therefore
an effective electroweak theory of massless fermions with finite momentum cut-off
and quartic Fermi interaction.
It is not indeed restrictive to consider only 
a quartic interaction, as 
after the first Renormalization Group (RG) integration step, monomials of all orders are generated in the interaction. 
One can even look to such theory as
the generic result of
integrating out high energy degrees of freedom from some more fundamental theory, see e.g. \cite{Pol}-\cite{Wi}. We prove that the fermionic functional integrals defining such theory are mathematically well defined 
and the correlations are {\it analytic} in 
a finite domain of couplings of order of the cut-off divided by the gauge boson masses. 
Analyticity follows from determinant bounds for fermionic expectations. The results holds in the limit of removed infrared cut-off,  so that the functional integrals are infinite-dimensional.


The Ward Identities in the effective theory have extra terms which would be formally vanishing removing the cut-off.
Keeping only classical gauge fields not taking into account interactions
the quadratic response 
is again sum of regularized triangle graphs which cancel out under the cancellation condition up to
terms of the order of the energy/cut-off ratio. The non perfect cancellation is expected as
finite momentum cut-off breaks gauge invariance already in the classical action. When the 
interaction is taken into account the response is sum of infinitely many graphs and the
Adler-Bardeen argument does not hold, as it relies on symmetries which are broken by the cut-off. However by
our exact RG analysis we show that the response can be decomposed in two terms;
one is
sum of triangle graphs with dressed vertices and propagators, which cancel out under the cancellation condition, and the other is given
by a complicate series of renormalized graphs which
again can be rigorously bounded
by a power of the energy/cut-off ratio.

In conclusion, the requirement that the current is conserved up to terms
of the energy divided by the cut-off scale, which is the natural condition in the effective QFT
as gauge symmetry is emerging,
provides the same constraint on charges as is found in the standard model at a perturbative level.  It is therefore 
a robust condition holding at a non-perturbative level in an effective theory and even when gauge symmetry is classically not exact. 

The results are obtained by Constructive RG methods, see e.g. \cite{M1} for an introduction,
which have been previously used to construct chiral interacting theories and the chiral anomaly 
in $d=1+1$ \cite{M11}-\cite{BFM2} in the limit of the removed ultraviolet cut-off.   
In the present case of effective electroweak in $d=3+1$ analyticity is found only
with a finite cut-off of the order of the gauge mass, and removed infrared cut-off,
as Fermi interaction is not renormalizable.
It would of course be interesting to get similar results for larger values of cut-off;
this could be obtained avoiding of integrating out the bosons and considering the standard model
electroweak theory with a momentum cut-off.  Going beyond perturbation theory is in this case
much more difficult as perturbation theory is expected to be non convergent and cluster expansions methods are needed.
Another interesting question is considering lattice regularization such that chiral symmetry is non broken at a classical level. In the case of QED, the perfect validity of
the Adler-Bardeen proved has been recently rigorously established \cite{GMP},\cite{GMP1}
 with a finite lattice (with emerging Lorentz symmetry), using the lattice regularization in \cite{NN}, and it would be interesting to extend such result, if possible,
to chiral theories.

The rest of the paper is organized in the following way. In \S II we present the effective model and we state our main result. In \S III we present the Renormalization Group analysis, in \S IV we establish analyticity and in \S V we derive the cancellation condition in the effective theory.

\section{Effective electroweak theory}

\subsection{Grassmann integration and currents}

We consider a single family of particles with 
two leptons, $(\n,e)$ and two quarks $(u,d)$;
the quarks have another color index which takes 3 values.
We introduce therefore 
Grassmann variables 
$\psi^\pm_{i,L,x},\psi^\pm_{i,R,x}$ 
with $L,R$ denoting chirality, 
$x\in [-L/2,L/2]^4$
with anti-periodic boundary conditions
and
\be
\{\psi^+_{i,s,x},\psi^+_{i',s',x'}\}=\{\psi^+_{i,s,x},\psi^-_{i',s',x'}\}=\{\psi^-_{i,s,x},\psi^-_{i',s',x'}\}=0\ee
with $s=L,R$. One introduces the doublets
$\Psi^\pm_{l,x}=(\psi^\pm_{\n,L,x},\psi^\pm_{e,L,x})$ and  $\Psi^\pm_{q,x}=(\psi^\pm_{u,L,x},\psi^\pm_{d,L,x})$.
The index labeling the 2 components of $\psi^\pm_{x,i,s}$ and the color index for  
$\psi^\pm_{u,s,x},\psi^\pm_{d,s,x}$
are omitted. 


We define $\psi^\pm_{i,s,x}={1\over L^4}\sum_k e^{i k x} \hat\psi^\pm_{i,s,k}$
with $k={2\pi\over L} n$, with $\hat \psi^\pm_{k,i,s}$ another set of Grassmann variables.
We introduce a smooth momentum cut-off $\chi_N(k)$ which is a infinitely differentiable compact support function (this is useful to get good decay properties in coordinate space)
such that $\chi_N(k)=0$ for $|k|\ge \g^{N+1}$ and 
$\chi_N(k)=1$ for $|k|\le  \g^{N}$, $\g>1$ a scaling parameter. Therefore $\g^N$ is the ultraviolet cut-off while the infrared cut-off is provided by $L$.
The
"fermionic gaussian measure" is defined as, $i=\n,e,u,d$, $s=L,R$
\be
P(d\psi)=[\prod^*_{i,s,k} d\hat\psi^+_{i,s,k}  d\hat\psi_{i,s,k}] 
e^{-{1\over L^4}\sum^*_k  \hat\psi^+_{i,s,k} \chi^{-1}_N(k) (i\s_\m^i k_\m)\hat\psi^-_{i,s,k}}
\ee
where $\prod^*_{k}$ is a product over $k$ in the support of $\chi_N(k)$ and
$\s_\m^L=(\s_0,i \vec \s) $ e $\s_\m^R=(\s_0,-i \vec \s) $, $\vec \s=\s_1,\s_2,\s_3$
with
\be
\s
_1=\begin{pmatrix}&0&1\\
          &1&0\end{pmatrix}
 \quad
\s_2=\begin{pmatrix}
&0&-i\\ &i&0
\end{pmatrix}
\quad\s_3=\begin{pmatrix}&1&0\\
          &0&-1\end{pmatrix}\ee 
The 2-point function is given by
\be
< \psi^-_{i,s,x}\psi^+_{i',s',y}>_0={\int P(d\psi) \psi^-_{i,s,x}\psi^+_{i',s',y}
\over\int P(d\psi)}=\d_{i,i'}\d_{s,s'}g_s(x,y)
\ee
with
\be
g_{i,s}(x,y)={1\over L^4} \sum_k e^{i k(x-y)}{\chi_N(k)\over -i \s^s_\m k_\m}
\ee
The $n$-point function $<\psi^{\e_1}_{l_1,s_1,x_1}...\psi^{\e_n}_{l_n,s_n,x_n}>_0$ is given by the Wick rule.
The cut-off function plays a very important role, as it makes the number of Grassmann variables finite, hence the Grassmann integral is well defined; at the end the limit $L\to\io$ is taken. 

The currents relevant
for electroweak interaction are the $W$ and $B$ ones
\be
j^k_{W,\m,x}={1\over 2} (j^k_{W,l,\m,x}+j^k_{W,q,\m,x})  \quad j_{B,\m,x}={1\over 2}
\sum_{i=\n,e,u,d\atop s=L,R }
( Y^L_i j_{i,\m,x}^L+ Y^R_i  j_{i
,\m,x}^R)\label{sap1}
\ee
where  $Y^L_i,  Y^R_i$ are the {\it hypercharges}, $Y^L_\n=Y^L_e$,  $Y^L_u=Y^L_d$ and
\be
j_{i,\m,x}^s= \psi^+_{i,s,x}\s_\m^s\psi^-_{i,s,x}\quad\quad
j^k_{W,\m,l,x}=\Psi^+_{l,x}\t^k \s_\m^L \Psi_{l,x}^-\quad\quad j^k_{W,\m,q,x}=\Psi^+_{q,x}\t^k\s_\m^L \Psi_{q,x}^-\label{7}
\ee
with $\t^k$ Pauli matrices. If $\psi$ are classical fields verifying the Dirac equation, the currents $j_{i,\m,x}^s, j^k_{W,\m,l,x}, j^k_{W,\m,q,x}
$
\pref{7} are separately conserved.
%
%

The form of the
interaction with classical fields is dictated by the requirement of invariance with respect to a gauge transformation; one gets that the average of the observable $O$, a monomial in the Grassmann variables,
is given by 
\be
<O>_{W,B}={\int P(d\psi) e^{\int dx( g W_{\m,x}^k j^k_{W,\m,x}+g' B_{\m,x} j_{B,\m,x})} O
\over \int P(d\psi) e^{\int dx(g W_{\m,x}^k j^k_{W,\m,x}+g' B_{\m,x} j_{B,\m,x})}}\label{ac1}
\ee
Note that such invariance is true at a classical level (that is formally replacing the Grassmann 
variables with functions) only in the limit of removed cut-off $N\to\io$ but is violated at finite $N$. 
Moreover, even when $N\to\io$ limit symmetry may be broken at a quantum level in
the functional integral \pref{ac}, what is exactly the anomaly phenomenon.

In order to see this, we can consider the average of the current with respect to \pref{ac1}.
It is computed by expanding in series in the gauge fields; 
if $A^0=B, A^1=W$ 
and the generating function is 
\be
e^{\WW^0_{W,B}}=  \int P(d\psi) e^{\int dx( g W_{\m,x}^k j_{W,\m,x}+ g' B_{\m,x} j_{B,\m,x})}\label{gem}
\ee
then, if the sum over the choices of $A$ and the combinatorial factor are understood, $g^0=g, g^1=g'$, $\e=0,1$ 
\be
<\hat  j_{B,\m,p}>_{W,B}
=\sum_n {1\over n!}\int dp_1... dp_n{\partial^{n+1} \WW^0_{W,B}\over \partial 
B_{\m,p}\partial 
A_{\m_1,p_1}^{\e_1}...\partial A^{\e_n}_{\m_n,p_n}}|_0  g^{\e_1} A^{\e_1}_{\m_1,p_1}..g^{\e_n} A^{\e_n}_{\m_n,p_n}\d(p+\sum_i p_i)\label{aa112}
\ee
Note that  the coefficients are simply the truncated correlations of currents
\be
{\partial^{n+1} \WW^0_{W,B}\over \partial 
B_{\m,p}\partial 
A^{\e_1}_{\m_1,p_1}...\partial A^{\e_n}_{\m_n,p_n}}|_0=
<\hat j_{B,\m,p};\hat j_{A^{\e_1},\m_1,p_1};..;\hat j_{A^{\e_n}, \m_n,p_n}>_0
\ee
so that
%
the expansion can be represented as sum of simple Feynman graphs, see Fig.1.
\insertplot{580}{79}
{\ins{50pt}{40pt}{$+$}
\ins{130pt}{40pt}{$+$}
}
{figff}
{\label{h2} Graphs contributing to \pref{aa112}
} {0}
We are in particular interested on the conservation of the $U(1)$ currents, which is expressed by
$<\partial_\m  j_{B,\m,x}>$.

\subsection{Ward Identities}

We introduce currents
$\hat j_{\m,1,p}=
\hat j^L_{\m,\n,p}+\hat j^L_{\m,e,p}$, $\hat j_{\m,2,p}=\hat j^L_{\m,u, p}+
j^L_{\m,d,p}$, $\hat j_{p,3,\m}=\hat j_{\m,i,p}^R$ $\a=1,2,3$ and a generating function
$\WW^0(A, \phi)$ in which we add to the exponent of \pref{gem}
a fermionic source term
$\sum_{\a=1,2,3} \int dx(\psi^+_{\a,x}\phi^-_{\a,x}+\psi^-_{\a,x}\phi^+_{\a,x})$,
where $\psi_1=\psi_{L,e}+\psi_{L,\n}$, 
$\psi_2=\psi_{L,u}+\psi_{L,d}$,  $\psi_3=\psi_R$. 
Conservation laws are encoded in Ward Identities, which can be derived performing in the generating function $\WW_0(A, \phi)$
the change of variables, 
$i=\n,e,u,d$ 
\be 
\psi_{i,L,x}^\pm\to e^{\pm i \a_{L,i,x}}  \psi^\pm_{i,L,x}\quad\psi^\pm_{i,R,x}\to e^{\pm i \a_{R,i,x}}  \psi^\pm_{i,R,x}
\quad \a_{L,\n,x}=\a_{L,e,x}\quad \a_{L,u,x}=\a_{L,d,x}
\label{gg1}
\ee
so obtaining, noting that the external currents are invariant $e^{\WW_0(A, \phi)}=$
\be
\int P(d\psi) e^{-\int dx  \psi^+_{s,i,x} (e^{i\a_{s,i,x}}D e^{-i\a_{s,i,x}}-D)\psi^-_{s,i,x} }
e^{\int dx( g W_{\m,x}^k j_{W,\m,x}+ g' B_{\m,x} j_{B,\m,x})+\int dx
(\psi_{i,s,x} e^{i\a_{i,s,x}} \phi_{i,s,x}+\psi_{i,s,x} e^{-i\a_{i,s,x}} 
\phi^+_{i,s,x })}\label{pip}
\ee
where
\be
D\psi_{s,i,x}=\int dk e^{-i k x} \chi^{-1}(k)\s^s_\m k_\m \hat\psi_{s,i,k}
\ee
The fermionic source term acquire a phase but not the current source. 
The Jacobian of the transformation is unitary as a straightforward consequence of the fact that the number of Grassmann variables $\hat\psi_k$ is 
{\it finite}; this is an important difference with respect to what happens in (formal) functional integrals with infinitely many variables \cite{F}. The exponent 
of the fermionic "measure" gets an extra term of the form
\be
\int dx \psi^+_{s,i,x} (e^{i\a_{s,i,x}}D e^{-i\a_{s,i,x}}-D)\psi^-_{s,i,x}=\sum_{i,s}
\int dx \d T_{i,s,x} \a_{i,s,x}
+O(\a^2) \label{lak}
\ee
where
\be
\d T_{s,i,x}={1\over L^8}\sum_{k,p}
 e^{-i p x}  \hat \psi^+_{s,i,k}
 \s^s_\m(\chi^{-1}(k) k_\m-\chi^{-1}(k+p)(k_\m+p_\m))\hat\psi^-_{s,i,k+p}\label{lep}
\ee
Replacing the cut-off function with $1$ one gets that $\chi^{-1}(k) k_\m-\chi^{-1}(k+p)(k_\m+p_\m)=p_\m$ so that the r.h.s. of \pref{lak} reduces to 
$\sum_{i,s}
\int dx j_{i,s,x} \partial_\m \a_{i,s,x}$. Note that the expressions $\chi^{-1}$ does not give any problem as identities have to be understood between correlations.
By performing derivatives of \pref{pip} with respect to $\a_{s,i,x}$ and to the external fields we get
\be
p_\m  <\hat j_{\m,\a,p};\hat\psi^-_{\a',k+p}\hat\psi^+_{\a',k}>+ 
<\d \hat j_{\a,p};\hat\psi_{\a',k+p}^+\hat\psi^+_{\a',k}>
=\d_{\a,\a'}(  <\hat\psi^-_{\a', k+p}\hat\psi^+_{\a',k+p}>-<\hat\psi^-_{\a',k} 
\hat\psi^+_{\a',k}>)\label{gg2}
\ee
with $\d \hat j_{1,p}=
\d\hat j^L_{\m,\n,p}+\d\hat j^L_{\m,e,p}$, $\d \hat j_{2,p}=
\d\hat j^L_{\m,u,p}+\d\hat j^L_{\m,d,p}$, $\d \hat j_{3,p}=
\d j_{\m,i,p}^R$ and
\be
\d \hat j_{\m,i,p}^s={1\over L^4}\sum_k  C^s(k,p) \hat\psi^+_{s,i,k}\hat\psi^-_{s,i,k+p}
 \quad\quad
C^s(k,p)=[(\chi^{-1}(k)-1) k_\m-(\chi^{-1}(k+p)-1)(k_\m+p_\m)]\s_\m^s\label{ggg}
\ee
The above identity has been written in a form closer to the formal WI writing $\chi^{-1}$
as $(\chi^{-1}-1)+1$ in \pref{lep}. 
With respect to the formal WI, \pref {gg2} has an extra term dependent on the momentum cut-off. The origin of ths term can be also understood from the equality between propagators
with momentum cut-off
\be
g_s(k)-g_s(k+p)=g_s(k)\s^s_\m p_\m g(k+p)+g_s(k) C(k,p) g_s(k+p)\label{che}
\ee
which replace the identity $g_s(k)-g_s(k+p)=g_s(k)\s^s_\m p_\m g(k+p)$
in presence of a cut-off.
%
%
%
%
%
\insertplot{580}{79}
{\ins{52pt}{40pt}{$=$}\ins{124pt}{40pt}{$-$}\ins{190pt}{40pt}{$+$}
}
{figjsp44b2}
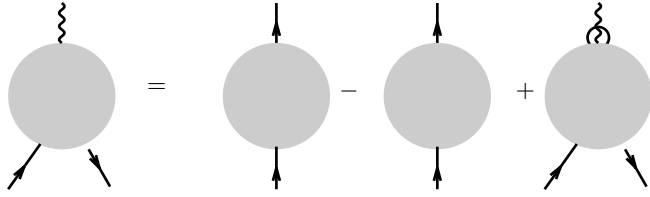
{\label{h2} The WI for the vertex function of \pref{gg2}; the last term is the vertex involving $\d j_\m$.
} {0}

We derive now the WI with respect to the currents; by performing derivatives in \pref{pip}
with respect to $\a$ and $W, B$ we get
\be
p_\m  <j_{\m,\a,p};\hat j_{A^{\e_1}, \m_1, p_1};..;\hat j_{A^{\e_n}, \m_n,p_n}
>=  <\d j_{p,\a};
\hat j_{A^{\e_1},\m_1,p_1};..;\hat j_{A^{\e_n}, \m_n,p_n}>\label{40}
\ee
Again we get an extra term with respect to the formal  WI proportional to $\d j_{p,\a}$;
if such term would be vanishing then from \pref{aa112}
we get the current conservation
$p_\m<\hat  j_{B,\m,p}>_{W,B}=0$.
Even more, the WI \pref{40} with the l.h.s. vanishing is equivalent 
to the separate conservation of currents of different species. The l.h.s. of \pref{40} is however
in general non vanishing.

\subsection{Cancellation condition}

Let us consider now the r.h.s. of \pref{40} with $n=2$ 
which is given 
\bea
&&p_\m <\hat j_{s,i,\m.p}; \hat j_{s,i,\n,p_1}; \hat j_{{s,i,\r,p_2}}>\\
&&=\int {dk \over (2\pi)^4}
tr {\chi(k)\over -i \s^s k_\m} C(k,p){\chi(k+p)\over -i \s_\m^s (k_\m+p_\m)}(-i \s^s_\n){\chi(k+p^2)\over -i \s^s_\m (k_\m+p^2_\m)}(-i \s^s_\r)+[(\n, p_1)\to (\r,p_2)]
\eea
and, see \cite{GMP}
\be
p_\m <\hat j_{s,p,\m}; \hat j_{s,p_1,\n}; \hat j_{s,p_2,\r}>=\e_s {1\over 12\pi^2}\e_{\n,\r,\a\,b}p^1_a p^2_\b+O(|\bar p|^3/\g^N)\label{21}
\ee
where $|\bar p|=\max(|p_1|,|p_2|)$ and $\e_L=-\e_R=1$. From \pref{21} we see that, in addition to terms proportional to the inverse cut-off, there are N-independent contributions 
which are the anomalies in the limit of removed cut-off.

The average of the $B$ current at second order, see \pref{aa112}, is given by
\be
<\hat j_{B,\m,p} ;\hat j_{W, \n, p_1}; \hat j_{W,\m, \r, p_2}>=
\hat L^W_{\m,\n,\r} (p_1,p_2)\quad\quad <\hat j_{B,\m,p}; \hat j_{B, \n, p_1}; \hat j_{B,\m, \r, p_2}>=\hat L^B_{\m,\n,\r} (p_1,p_2)\ee
The divergence follows by
\pref{21} and one finds
\bea
&&p_\m \hat L^W_{\m,\n,\r} (p_1,p_2)=
{1\over 12\pi^2}\e_{\n,\r,\a,\b}p^1_\a p^2_\b [\sum_i Y^L_i]+R^W_{\n,\r}(p_1,p_2)
\nn\\
&&p_\m \hat L^B_{\m,\n,\r} (p_1,p_2)=
{1\over 12\pi^2} \e_{\n,\r,\a,\b} p^1_\a p^2_\b [
\sum_i (Y^L_i)^3-(Y^R_i)^3]+R^B_{\n,\r}(p_1,p_2)
\label{sper}
\eea
with 
\be
|R^W_{\n,\r}(p_1,p_2)|, |R^B_{\n,\r}(p_1,p_2)|\le C {|\bar p|^3\over \g^N}\label{sper1}
\ee
If we require that the current is conserved up to terms  to 
the energy divided by the cut-off scale we get
\be
\sum_i Y^L_i=0\quad\quad \sum_i (Y^L_i)^3-(Y^R_i)^3=0\label{79}
\ee
These are of course the same conditions found
in the standard electroweak theory with classical gauge fields; indeed the limit $N\to\io$ can be taken safely and exact conservation is found. The condition is verified by elementary particles as 
$Y^L_i$ has value $(-1, -1,{1\over 3},{1\over 3})$, and $Y^R$ has value $(0,-2,{4\over 3}, -{2\over 3})$ so that one gets
$-2+6{1\over 3}=0$ and $6(1/3)^3+2(-1)^3-3(4/3)^3-3(-2/3)^3-(-2)^3=0$.
The conservation of $W$ current does not give further constraint.

We want to investigate if the condition \pref{79} still ensures the current conservation in the interacting case. In such a case the terms contributing to the divergence of the current are a series of infinitely many Feynman graphs and a direct verification is impossible. In addition, in order to get non perturbative results one needs to keep a finite ultraviolet cut-off, as the Standard Model is not asymptotically free. We ask therefore if also in the interacting case
with finite cut-off
the current is conserved up to terms proportional to 
the energy divided by the cut-off scale provided that the condition \pref{79}
is true.

\subsection{Effective Fermi interaction}

The Standard electroweak theory is obtained replacing in \pref{gem} the field 
$B_\m,W_\m$ with the sum of two fields $B_\m+\tilde B_\m$ and $W_\m+\tilde W_\m$, 
, see e.g. \cite{T}, where $B_\m,W_\m$ are
 classical background filelds and $\tilde B,\tilde W$ are quantum fields, with a gauge invariant action. The $Z$ and $e.m.$ currents are define by
\be
\int dx ( g \tilde W^3_{x,L} j_{W,\m}^3+ g' \tilde B_x j_{B,\m})=\int dx (e A_{\m,x} j^{em}_{\m,x}+\bar g' Z_{\m,x} j^Z_{\m,x})
\ee
where $\bar g'={g\over \cos\th}$, $\tanh\th=g'/g$, $g\sin\th=g' \cos\th=e$ 
and the charges are
\be
2 Q^s_i=I_{3,i}^s+Y^s_i \label{ss} \ee 
with $I_{3,i}^L=\pm 1$ and $I_{3,i}^R=0$ (so that $Q_i$ is $(0,-1,2/3,-1/3)$) and 
\be
j^{e.m.}_{\m,x }=
e \sum_i Q_i (j_{i,\m,x}^L+j_{i,\m,x}^R)\quad\quad 
j_{Z,\m,x}=\sum_{i,s} ( I_{3,i}^s -\sin^2\th Q_i) j_{i,\m,x}^s
\ee
From \pref{ss} we see that the proof of charge quantization follows from the quantization
of the hypercharges, as $I_{3,i}^s$ is quantized.

Due to the Higgs mechanism, the quantum $\tilde Z_\m$ and $\tilde W$ gauge fields acquire a mass. The effective electroweak theory is obtained integrating the boson fields generating an effective quartic interaction; it is indeed 
not restrictive to consider only quartic interactions as monomials of any order in the fields are generated during the RG integrations, see \S 3. Neglecting for the moment the external gauge fields 
the correlations of the effective theory are given by
\be 
<O>=
{\int P(d\psi) e^{V(\psi)} O\over \int P(d\psi) e^{V(\psi)} } 
 \label{xx}  \ee
where $P(d\psi)$ is the fermionic integration with renormalized propagator 
\be
g_{i,s}(x,y)={1\over Z_{N,i,s}}
{1\over L^4} \sum_k e^{i k(x-y)}{\chi_N(k)\over -i \s^s_\m k_\m}
\ee
and
\be
V(\psi)=\int dx dy  \l [ W(x,y) (j_{W,\m,x}^1 j_{W,\m,x}^1+j_{W,\m,x}^2 
j_{W,\m,x}^2)+ w_Z(x,y) j_{Z,\m,x} j_{Z,\m,x} ]\label{sss}
\ee
with $\l$ an effective coupling proportional to $g^2$ and
%
%
\be
v_W(x,y) =\int dk e^{i k (x-y)}{ \chi_N(k)\over |k|^2+M_W^2}\quad\quad v_Z(x,y) =A\int dk e^{i k (x-y)}{ \chi_N(k)\over |k|^2+M_Z^2}
\ee
with $M=M_Z>M_W$ and $A$ is a constant to take into account the difference in the effective couplings and masses.

At finite $N$ we can prove that the this effective theory has a well definite non-perturbative meaning, even if in the $L\to\io$ limit the functional integrals are infinite dimensional.
Indeed
in \S 4 we prove the following result 
\vskip.3cm
{\bf Theorem 1} {\it The correlations corresponding to \pref{xx} are analytic in $\l$ for
$|\l|\le [{M\over C \g^{N} }]^6
$ uniformly as $L\to\io$
}
\vskip.2cm
Analyticity in the coupling around the origin is a remarkable fact due to the purely fermionic nature of \pref{xx}; indeed,
in presence of bosons analyticity in zero cannot be true due to 
Dyson argument. The estimated radius of convergence is proportional to the gauge mass divided by the cut-off; this reflects
the perturbative non-renormalizability of the theory, and implies that the cut-off must be chosen of the order of the gauge mass. The proof is based on non-pertubative methods and avoid the Feynman graph expansion, see \S III.

\subsection{Effective electroweak theory and main result}
 

We include in the effective model the external gauge fields associate to the $B$ and $W$ currents. Due to the interaction, the charges are renormalized and one needs to introduce bare currents depending on parameters to be fixed so that the ir values correspond to
the physical values at low momenta.
We introduce therefore the {\it bare} background currents 
\be
\tilde j^k_{W,\m,x}=\sum_{a=l,q}  Z^W_{N,a,k}  j^k_{W,a,\m,x}  \quad\quad \tilde j_{B,\m,x}=
\sum_{i=\n,e,u,d\atop s=L,R }
Y^s_i  Z^J_{N,i,s}  j_{i,\m,x}^{s} 
\label{sap11}
\ee
with the parameters $Z^W_{N,a,k}$ and $Z^J_{N,i,s}$ to be chosen in order to fix the dressed parameters, which can be obtained by the correlations. It is indeed an
outcome of our RG analysis in \S 2 that in the analyticity domain $|\l|\le [{M\over C \g^{N} }]^6
$ the 2-point function is
\be
<\hat\psi^+_{k,i,s} \hat\psi^-_{k,i,s} >={1\over Z_{-\io,i,s}} {1\over -i \s_\m^s k_\m}(1+R(k))\label{gam}
\ee
with $Z_{-\io,i,s}$ is a non trivial analytic function of $\l$ representing the
wave function renormalization and  $|R(k)|\le C |\l| |k| \g^{-N}$.
Similarly the 3-point functions are, $k\sim k+p\sim \k$
\be
<\tilde j_{B,\m,x};
\hat\psi^+_{i,s,k} \hat\psi^-_{i,s,k+p}>=
{1\over \s^{s}_\m k_\m}\s^{s}_\m{1\over \s^{s}_\m (k_\m+p_\m)}[ ( { Y_i^s Z_{i,s,-\io}^J\over 
Z_{i,s,-\io} Z_{i,s,-\io}}+R(k,k+p)]\label{lll}
\ee
with  $|R(k,k+p)|\le C |\l| |\k| \g^{-N}$
from which we see that the dressed hypercharge is ${ Y_i^s Z_{i,s,-\io}^J\over 
Z_{i,s,-\io}}$. A similar expression is found for $<\tilde j_{W,\m,x}
\hat\psi^+_{\n,s,k} \hat\psi^-_{e,s,k+p}>$ in which the dominant term is proportional
to  ${ Z_{a,-\io}^W\over 
Z_{\n,L,-\io} Z_{e,L,-\io}}$.

The bare normalization are chosen in order to ensure the following conditions
\be Z_{i,s,-\io}=1 \quad\quad  Z^J_{i,s,-\io}=1\quad\quad Z_{a,-\io}^W=1\label{lll1}
\ee
The first condition ensures that the wave renormalization in the low-energy limit is the same for all particles; the second that the dressed hypercharge is equal to $Y_i^s$
and the third that the normalizations in the $W$ currents do not depend on the particle species in the low energy limit. The non-trivial renormalization of the charges is related to the extra terms
with $\d j$ in the WI for the 3-point function \pref{gg2}.
Such WI holds also in the interacting case, as $V$ is invariant under the transformation 
\pref{gg1}. However the term depending on $\d j$, which is proportional to the inverse of the cut-off in the non interacting case, is $N$ independent up to small corrections and $O(\l)$ in presence of interaction, see \cite{BFM1},\cite{BFM2}  for a similar phenomenon in the $d=1+1$ case.

The effective electroweak theory replacing \pref{gem} is therefore given by
\be
<O>_{W,B}={\int P(d\psi) e^{V(\psi) + \int dx (g W_{\m,x}^k \tilde j^k_{W,\m,x}+g' B_{\m,x} \tilde j_{B,\m,x})} O
\over \int P(d\psi) e^{V(\psi)+\int dx( g W_{\m,x}^k \tilde j_{W,\m,x}+g' B_{\m,x} \tilde j_{B,\m,x})}}\label{ac}
\ee
with $V$ given by \pref{sss} and $\tilde j^k_{W,\m,x}$ are given by 
\pref{sap11}
 with the normalization condition \pref{lll} and \pref{lll1}.
The response of the $U(1)$ current in the effective theory is given by 
\be
<\hat  j_{B,\m,p}>_{W,B}
=\sum_n {1\over n!}\int dp_1... dp_n{\partial^{n+1} \WW_{W,B}\over \partial 
B_{\m,p}\partial 
A_{\m_1,p_1}^{\e_1}...\partial A^{\e_n}_{\m_n,p_n}}|_0  g^{\e_1} A^{\e_1}_{\m_1,p_1}..g^{\e_n} A^{\e_n}_{\m_n,p_n}\d(\sum_i p_i)\label{aa11}
\ee
with the derivative above given by 
$<j_{B,\m,p};j_{A^{\e_1},\m_1,p_1};..;j_{A^{\e_n}, \m_n,p_n}>
$
\insertplot{580}{79}
{\ins{50pt}{40pt}{$+$}
\ins{130pt}{40pt}{$+$}}
{figjsp44b}
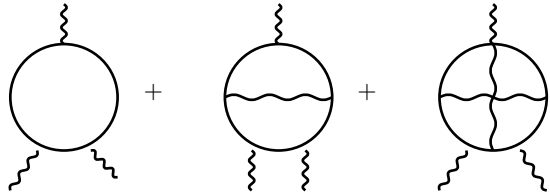
{\label{h2} Graphs contributing to the expansion at $n=2$.
} {0}
and 
\be 
e^{\WW(A) }=
\int P(d\psi) e^{V+\int dx( g W_{\m,x}^k \tilde j^k_{W,\m,x}+g' B_{\m,x} \tilde j_{B,\m,x})}\equiv 
\int P(d\psi) e^{V+\BB(A)} \label{sac} \ee 
There are now radiative corrections, see Fig 3,
which could produce extra conditions in order to impose that the current is conserved. This is however excluded by the following result.
\vskip.3cm {\bf Theorem 2}
{\it For $|\l|\le [{M\over C \g^{N} }]^6
$ and choosing $Z_{i,s,N}, Z^J_{i,s,N}, Z_{a,N}^W$ as functions of $\l$ so that
\pref{lll1} holds, then
the 3-point function can be written as 
\bea
&&<\tilde j_{B, \m,x};\tilde j_{B, \n,x_1};\tilde j_{B, \r,x_2}>=L^B_{\m,\n,\r}(x,x_1,x_2) +R^{1,B}_{\m,\n,\r}(x,x_1,x_2)\label{sec1}\\
&&<\tilde j_{B, \m,x};\tilde j_{W, \n,x_1};\tilde j_{W, \r,x_2}>=L^W_{\m,\n,\r}(x,x_1,x_2) 
+R^{1,W}_{\m,\n,\r}(x,x_1,x_2)\nn
\eea
with $\hat L^B_{\m,\n,\r}, \hat L^W_{\m,\n,\r}$ verifying \pref{sper} and
\be
|R^{1,B}_{\m,\n,\r}(x,x_1,x_2)|, |R^{1,W}_{\m,\n,\r}(x,x_1,x_2)|
\le C
 [{1\over \g^N \d}]^{1\over 2} C_\d \label{sec}
\ee
with $\d$ is the minimal distance between $x,x_1,x_2$, 
}
\vskip.3cm
We see from \pref{sec1} that also in the interacting case the current is conserved up
to terms proportional to the inverse of the cut-off scale, provided that the conditions
$
\sum_i Y^L_i=0$ and $\sum_i (Y^L_i)^3-(Y^R_i)^3=0$ hold; even if the average of the
current
is given 
by a complicate series of graphs, no new conditions arises. The crucial bound \pref{sec} is non-perturbative; graphs expansion is avoided and determinant bounds are used to implement cancellations due to Pauli principle and ensuring analyticity.

\section{Renormalization Group analysis}

The starting point for the analysis of $\WW_{B,W}(A)$ \pref{sac} 
is the following decomposition of the cut-off function
\be
\chi_N(k)=\sum_{h=-\io}^N f_h(k)\quad f_h(k)=\chi(\g^{-h} k)-\chi(\g^{-h+1} k)
\ee
so that $f_h(k)$ is a smooth cut-off function selecting momenta
in $\g^{h-1}\le |k|\le \g^{h+1}$; we also call 
$\chi_h(k)=\sum_{j=-\io}^h f_j(k)$ the cut-off function selecting momenta $|k|\le \g^h$. 
We perform an exact RG integration. The starting point is the addition property
\be
P(d\psi)=P(d\psi^{(N)})P(d\psi^{(\le N-1)})
\ee
where $P(d\psi^{(N)})$ and $P(d\psi^{(\le N-1)})$ is the gaussian grassmann measure with propagators 
\be
g^{(N)}_{i,s} (x,y) = {1\over Z_{N,i,s} }{1\over L^4}\sum_k e^{i k(x-y)} {f_N(k)\over -i \s_\m^s k_\m}\quad\quad
g^{(\le N-1)_{i,s} }(x,y)={1\over Z_{N,i,s} }{1\over L^4}\sum_k e^{i k(x-y)} {\chi_{N-1}(k)\over -i \s_\m^s k_\m}\ee
Setting $\VV^{(N)}=V+\BB$ we can write
\be
\int P(d\psi) e^{\VV^{(N)}}
=\int P(d\psi^{(N)})P(d\psi^{(\le N-1)}) e^{\VV^{(N)}}=\int P(d\psi^{(\le N-1)})
e^{\VV^{(N-1)}}
\ee
where by definition
\be
e^{\VV^{(N-1)}}=\sum_{n=0}^\io {1\over n!}\EE^T_n(\VV; n)
\ee
where $\EE^T_n(\VV; n)$ are the {\it truncated expectations} (that is the sum of connected Feynman graphs)
with propagator $g^{(N)}$. After the integration of the first scale $\psi^{(N)}$ we get an effective potential which is sum of infinitely many monomials in the  $\psi^{(\le N-1)},A$
\be
\VV^{(N-1)}(A, \psi^{(\le N-1)})=\sum_{l,m=0}^\io  \int d\underline x \; W^{(N-1)}_{\underline i, 
\underline s, l,m}(\underline x)\prod_{j=1}^l 
\psi^{\e_j, (\le N-1)}_ {i_j,s_j,x_j} \prod_{j=1}^m A_{\m_j, x_j}^{\e_j}\label{nm}
\ee
The scaling dimension is $D=4-{3\over 2}l-m$ and we can separate the irrelevant terms $D<0$ from the rest; moreover marginal and 
relevant terms $\psi^+\psi^-$ or  $A \psi^+\psi^-$  are generally non-local (that is the fields have different coordinates), and we can split them in a local plus an irrelevant part.
In order to obtain this we define a localization operator $\LL$ such that $\LL$ gives a vanishing result on the irrelevant terms and 
\bea
&&\LL \psi^{+, (\le N-1)}_ {i,s,x} \psi^{-, (\le N-1)}_ {i',s,y}=\psi^{+, (\le N-1)}_ {i,s,x} 
\psi^{-, (\le N-1)}_ {i',s,x}+(x-y)_\m\psi^{+, (\le N-1)}_ {i,s,x} 
\partial_\m \psi^{-, (\le N-1)}_ {i',s,x}\nn\\
&&\LL A_{z}^\e  \psi^{+, (\le N-1)}_ {i,s,x} \psi^{-, (\le N-1)}_ {i',s,y}=A^\e_{\m,z } \psi^{+, (\le N-1)}_ {i,s,z} 
\psi^{-, (\le N-1)}_ {i',s,z}
\eea
In momentum space the above expression can be equivalently written as
\bea
&&\LL \int dk \hat W_{2,0}(k)  \hat\psi^{+, (\le N-1)}_ {i,s,k} \hat\psi^{-, (\le N-1)}_ {i',s,k}=
\int dk (\hat W_{2,0)}(0)+k_\m \partial_\m \hat W_{2,0}(0))
  \psi^{+, (\le N-1)}_ {i,s,k} \psi^{-, (\le N-1)}_ {i',s,k}\nn\\
&&\LL \int dk dp \hat W_{2,1,\m}(k,p)
A_{\m,p}^\e  \hat\psi^{+, (\le N-1)}_ {i,s,k} \hat \psi^{-, (\le N-1)}_ {i',s,k+p}=\int dk dp
\hat W_{2,1,\m}(0,0)
A_{\m,p}^\e  \hat\psi^{+, (\le N-1)}_ {i,s,k} \hat \psi^{-, (\le N-1)}_ {i',s,k+p}
\eea
The fields $\psi$ have the same chirality, as the propagators are diagonal in the chiral index and the currents have the same chirality.
By parity of the propagator $\hat W_{2,0}(0)=0$.
Lorentz symmetry, valid also in presence of cut-off, implies that 
$\partial_\m \hat W_{2,0}(0)$ and $\hat W_{2,1,\m}(0,0)$ are proportional to $\s_\m^s$.
There are no contributions $\psi^+_i \partial \psi^-_{j}$ with $i\not= j$.
Indeed if $i$ and $j$ belongs to different families then such term would violate the invariance
under a global phase transformation $\Psi_a^\pm\to e^{\pm i \a_a}\Psi_a^\pm$
with $a=l,q$. If $i,j$ belong to the same family then if the field $i$ has $s=R$ is impossible by a similar argument; if $s=L$
we call $n_1$ the number of vertices containing
only one field $i$ (say $e$), $n_2$ or $n_4$
the number of vertices containing $2$ or $4$ fields; then
$(n_1-1+2 n_2+4n_4)/2$ must be integer; hence $n_1$ is odd
but then there is an odd number of fields of the other family $u$ or $d$ and this is impossible.
The marginal quadratic terms have therefore the form  $z_{N-1,i,s}\int dk k_\m \psi^+_{i,s,k} \s^s_\m \psi^+_{i,s,k} $ which can be included in the wave function renormalization
\be
Z_{N-1,i,s}=Z_{N,i,s} +z_{N-1,i,s} 
\ee
In the same way there are no contribution to $W_{2,1,\m}$ with fields with different $i$ index if the source is diagonal in the index, and
the non vanishing terms
can be included in the current renormalizations defining
\be
Z^J_{N-1,i,s}=Z^J_{N,i,s}+z^J_{N-1,i,s}\quad\quad  Z^W_{N-1,a,k}= Z^W_{N,a,k}+z^W_{N-1,a,k}
\ee
In conclusion we get
\be
\int P(d\psi^{(\le N-1)})
e^{\LL \VV^{(N-1)}+\RR \VV^{(N-1)}}
\ee
with propagator
\be
g^{(\le N-1)}(x,y)={1\over Z_{N-1,i,s} } {1\over L^4}\sum_k e^{i k(x-y)} {\chi_{N-1}(k)\over -i \s_\m^s k_\m}
\ee
where 
\be
\RR\VV^{(N-1)}(A, \psi^{(\le N-1)})=\sum^*_{l,m} \int d\underline x \;
\tilde W^{(N-1)}_ {n,m}(\underline x)\prod_{j=1}^l 
\partial^{s_i}\psi^{\e_j, (\le N-1)}_ {i_j,s_j,x_j} \prod_{j=1}^m A^{\e_j}_{\m_j,x_j}\label{nm}
\ee
where $\sum^*$ has the constraint that if $l=2, m=0$ then $s_1+s_2=2$ and if  $l=2, m=1$ then $s_1+s_2=1$; that is, the effect of the $\RR$ operation is to produce a series with negative scaling dimension. Moreover
\be
\tilde\LL\VV^{N-1)}= \int d x (g W_{\m,x}^k \tilde j^{k,(N-1)}_{W,\m,x}+g' B_{\m,x} \tilde j^{(N-1)}_{B,\m,x})
\ee
with
\be
\tilde j^{k,(N-1)}_{W,\m,x}=\sum_{a=l,q}  Z^W_{N-1,a,k}  j^{k,(N-1)}_{W,a,\m,x}  \quad\quad   \tilde j^{(N-1)}_{B,\m,x}=
\sum_{i=\n,e,u,d\atop s=L,R }
Y^s_i  Z^J_{N-1,i,s}  j_{i,\m,x}^{s,} (N-1)
\label{sap1}
\ee
The expression is similar to the initial one, with the difference that the effective potential is sum over monomials of all orders and with derivatives in the fields, and 
the wave function renormalizations and the normalizations of the charges is modified. 
The field $\psi^{(N-1)}$ can be integrated and the procedure can be iterated.
The generic RG integration step gives
\be
e^{W(A)}=
\int P(d\psi^{(\le h)}) e^{\LL\VV^h+\RR\VV^h }\label{xx1}  
\ee
where $P(d\psi^{(\le h)})$ is a Grassmann integration
with propagator 
\be
g^{(\le h))}_{i,s}
= {1\over Z_{i,s,h}}{\chi_h(k)\over -i \s^i_\m k_\m}
\ee
and $\LL\VV^h,\RR\VV^h$ are similar with $N-1$ replaced by $h$.

As an outcome of the above construction 
the kernels $W^{(h)}_ {n,m}$ are expressed by convergent series in $\l$
\be
W^h_{l,m}=\sum_{n=1}^\io K^h_{n, l,m} \l^n
\ee
and in section \S III it is proved that
\be  
|K^h_{n, l,m}|\le C^{l+n+m}  \g^{(4-(3/2)l-m) h} 
\g^{ \d_n(h-N)} 
[{\g^{6N}\over M^6 }
]^n
\label{bb}
\ee
with $\d_0=0$ and $\d_n=\th=1/2$ for $n\not=0$. The factor $\g^{ \th (h-N)}$ is a gain with respect to the "dimensional bound" in the term with at least a $\l$, and is due to the dimensional irrelevance of the quartic terms; such extra factor plays a crucial role in the following.
Note that the estimated convergence radius is proportional to the cut-off and mass ratio, as a consequence of the perturbative non-renormalizability of the theory.

The effective renormalizations verify recursive equations, if $\ZZ_h=(Z_{i,s,h},Z^J_{i,s,h},
Z^W_{a,h})$
\be
\ZZ_{h-1}=\ZZ_h+\b^h_\ZZ(\l;\ZZ_h,..,\ZZ_N)
\quad\quad |\b^h_\ZZ|\le \g^{ \th (h-N)} C  [{\g^{6N}\over M^6 }\l]\label{gamma1}
\ee
where the  r.h.s. have an extra factor $\g^{ \th (h-N)}$ 
by \pref{bb}, noting that there is no contribution to the $\b$ function of order zero. The renormalizations are therefore finite
\be
\ZZ_{h-1}=\ZZ_{N}+\sum_{k=h}^N \b^k_\ZZ(\l;\ZZ_k,..,\ZZ_N)\label{sep} 
\ee
We impose the renormalization conditions;  we can look to \pref{sep} as a self-consistence equation and by a contraction methods we find $\ZZ_{i,s,N}$
as a function of $\l$ so that \be \ZZ_{h}=1+O(\g^{\th(h-N)} \l {\g^{6N}\over M^6})\label{sep1} 
\ee 
Analyticity stated  in
Theorem 1 is an immediate consequence of \pref{bb}. Note that the denominator of the correlations (the partition function)
at finite $L$ is analytic for any $\l$ in the whole complex plane at finite $L$ as it is a finite dimensional Grasmann integral; on the other hand the RG analysis above provides an expansion
which coincides order by order and is analytic in a finite domain, so that it fully reconstructs the partition function.
The correlation is also analytic, as the denominator is non vanishing in a finite disk for small $\l$
for any $L$
and the numerator is a finite dimensional integral; moreover
it coincides order by order with the expansion found analyzing the generating function by RG
which is also analytic in the same domain so that they coincide and analyticity as $L\to\io$ follows.

%
%
\section{Convergence and analyticity}

We prove now the bound \pref{bb}.
The kernels of the effective potential generated in the Renormalization group analysis 
can be conveniently written as a sum of {\it trees}, defined in the following way, see e.g. \cite{M1}.

\insertplot{400}{200}
{\ins{60pt}{90pt}{$v_0$}\ins{120pt}{100pt}{$v$}
\ins{100pt}{90pt}{$v'$}
\ins{120pt}{-5pt}{$h_v$}
\ins{235pt}{-5pt}{$N$}
\ins{255pt}{-5pt}{$N+1$}
}
{treelut2}{\label{n11} A labeled tree 
}{0}


Let us consider the family of all trees which can be constructed
by joining a point $r$, the {\it root}, with an ordered set of $n\ge 1$
points, the {\it endpoints} of the {\it unlabeled tree}, 
so that $r$ is not a branching point. $n$ will be called the
{\it order} of the unlabeled tree and the branching points will be called
the {\it non trivial vertices}.
The unlabeled trees are partially ordered from the root to the endpoints in
the natural way; we shall use the symbol $<$ to denote the partial order. 
The number of unlabeled trees is $4^n$.
The set of labeled trees $\TT_{h,n}$
is defined associating a label $h\le N-1$ with the root; 
moreover  
we introduce
a family of vertical lines, labeled by an an integer taking values
in $[h,N+1]$ intersecting all the non-trivial vertices, the endpoints and other points called trivial vertices.
The set of the {\it
vertices} $v$ of $\t$ will be the union of the endpoints, the trivial vertices
and the non trivial vertices. The scale label is $h_v$ and, if $v_1$ and $v_2$ are two vertices and $v_1<v_2$, then
$h_{v_1}<h_{v_2}$.
Moreover, there is only one vertex immediately following
the root, which will be denoted $v_0$ and can not be an endpoint;
its scale is $h+1$. The end-points are associated $V(\psi^{(\le N)})$ , and in such a case
the scale is $N+1$ and are named as normal-end-points, or a source terms $\BB(\psi^{(\le N)},A)$ or
$\LL\VV^{h_v-1}(\psi^{(\le h_v-1)},A)$ and in this case the scale is $h_v \le N+1 $ 
and there is the constraint that
that $h_v=h_{v'}+1$, if $v'$ is the first non trivial vertex immediately preceding $v$;
in such a case they are called special end-points.

The effective potential can be written as 
\be
\VV^{(h)}(\psi^{(\le h)},A) =
\sum_{n=1}^\io\sum_{\t\in\TT_{h,n}}
\VV^{(h)}(\t)\;,\ee
where, if $v_0$ is the first vertex of $\t$ 
and $\t_1,..,\t_s$ ($s=s_{v_0}$)
are the subtrees of $\t$ with root $v_0$,
$\VV^{(h)}$
is defined inductively by the relation, $h\le N-1$
$$\VV^{(h)}(\t)={(-1)^{s+1}\over s!} \EE^T_{h+1}[\bar
\VV^{(h+1)}(\t_1);..; \bar
\VV^{(h+1)}(\t_{s})]$$
where $\EE^T_{h+1}$ is the {\it truncated expectation}
and $\bar
\VV^{(h+1)}(\t)=\RR \VV^{(h+1)}(\t)$ if
the subtree $\t_i$ contains more then one end-point,
while if $\t_i$ contains only one end-point $\bar
\VV^{(h+1)}(\t)$ is
$V(\psi^{(\le N)})$ if is a normal end-point (and in such case $h=N-1$)  
or if is a special end-point
$\LL\VV^{h+1}(A, \psi^{(\le h+1)})$, $h<N-1$ or $\BB(\psi^{(\le N)},A)$.
We define
$P_v$ as the set of field labels of $v$
representing the external fields 
and if
$v_1,\ldots,v_{s_v}$ are the $s_v$
vertices immediately following $v$, then we denote by $Q_{v_i}$ the intersection of $P_v$ and
$P_{v_i}$; this definition implies that $P_v=\cup_i Q_{v_i}$. The
union of the subsets $P_{v_i}\bs Q_{v_i}$ are the internal fields of $v$.
Therefore if $\PP_\t$ is the familiy of all such choices and $\bP$
an element we can write
\be
 \VV^{(h)}(\t)=\sum_{\bP\in\PP_\t}
\int dx_{v_0} W^{(h+1)}_{\t,\bP}(x_{v_0})[\prod_{f\in P_{v_0}}
\psi^{\e(f)(\le h)}_{x(f)}][\prod_f A(x_f)]
\ee
where $W^{(h_v)}_{\t,\bP}(x_{v_0})$ is defined inductively by the equation
\be
W^{(h+1)}_{\t,\bP}(x_{v})={1\over s_{v} !} [\prod_{i=1}^{s_{v_i}}W^{(h_v+1)}_{\t,\bP}(x_{v_i})] \EE^T_{h_v}(
\tilde\psi^{(h_v)}(P_{v_1}/Q_{v_1});
...;\tilde\psi^{(h)_v}(P_{v_{s_v} }/Q_{v_{s_v}})
\ee
where $\tilde\psi^{(h)}(P)=\prod_{f\in P}\psi^{(h)\e(f)}_{x(f)}$ and $x_v$ are the coordinates associated to the vertex $v$.
We use the following well known representation of the fermionic
truncated expectation,, if $P$ is a set of indices
\be
\EE^T_{h}(\tilde\psi^{(h)}(P_1);\tilde\psi^{(h)}(P_2);
...;\tilde\psi^{(h)}(P_s))= \sum_{T}\prod_{l\in T} g^{(h)}(x_l-y_l)
\int dP_{T}(\tt) \det
G^{h,T}(\tt)\ee
where $T$ is a set of lines forming an {\it anchored tree graph} between
the clusters of points $x(f)_{f\in P_i}$, that is $T$ is a
set of lines, which becomes a tree graph if one identifies all the
points in the same cluster. Moreover $\tt=\{t_{i,i'}\in [0,1],
1\le i,i' \le s\}$, $dP_{T}(\tt)$ is a probability measure with
support on a set of $\tt$ such that $t_{i,i'}=\uu_i\cdot\uu_{i'}$
for some family of vectors $\uu_i\in \RRR^s$ of unit norm. Finally
$G^{h,T}(\tt)$ is a $(n-s+1)\times (n-s+1)$ matrix, whose elements
are given by
$G^{h,T}_{ij,i'j'}=t_{i,i'} 
g^{(h)}(x_{ij}-y_{i'j'})$.

By inserting the above representation we can write
$
W^{(h+1)}_{\t,\bP}=\sum_\bT W^{(h+1)}_{\t,\bP,\bT}
$
where $\bT$ is the union of all the trees $T$.

The determinants are bounded by the {\it Gram-Hadamard inequality}, stating
that, if $M$ is a square matrix with elements $M_{ij}$ of the form
$M_{ij}=<A_i,B_j>$, where $A_i$, $B_j$ are vectors in a Hilbert space with
scalar product $<\cdot,\cdot>$, then
$$|\det M|\le \prod_i ||A_i||\cdot ||B_i||\;$$
where $||\cdot||$ is the norm induced by the scalar product.
Let $\HH=\RRR^s\otimes \HH_0$, where $\HH_0$ is the Hilbert space of complex
two dimensional vectors with scalar product
$<F,G>=\int dk F^*_i(k) G_i(k)$.
It is easy to verify that
\be
G^{h_v,T_v}_{ij,i'j'}=t_{i,i'} g^{(h_v)}(x_{ij}-y_{i'j'})
=<\uu_i\otimes A^{(h_v)}_{x(f^-_{ij})
},
\uu_{i'}\otimes B^{(h_v)}_{x(f^+_{i'j'})}>\ee
where $\uu_i\in \RRR^s$, $i=1,\ldots,s$, are the vectors such that
$t_{i,i'}=\uu_i\cdot\uu_{i'}$ and $A, B$ suitable functions. 
The integrals over the coordinate are done integrating over the tree $T$ and the interactions, using that for any $K$
\be
|v(x)|\le {\g^{4 N}\over M^2 } {C_N\over 1+(M |x|)^K}\quad\quad \int dx |v(x)|\le  {\g^{4 N}\over M^6 }
\ee
and
\be
|g^h(x)|\le C \g^{3 h} e^{-(\g^h |x|)^{1\over 2}} \quad\quad \int dx |g^h(x)|\le C  \g^{- h} 
\ee
In conclusion we get
$$\int d x_{v_0} |W_{\t,\bP,T}(x_{v_0})|\le L^4 \prod_v {1\over s_v!} 
C^{\sum_{i=1}^{s_v}|P_{v_i}|-|P_v|} \g^{-4 h_v (s_v-1)} 
\g^{3/2 h_v (\sum_i |P_{v_i}|-|P_v|)}[\prod_v \g^{-z_v } 
[{\g^{4 N}\over M^6 }]^n$$
where $z_v=1$ if  the external fields are $\psi\psi$ or $A\psi\psi$ and zero otherwise.
By using that ( $h_v-h_{v'}=1$)
\bea
&&\sum_v (h_v-h)(s_v-1)=\sum_v (h_v-h_{v'})(m^4_v+n^A_v-1)\nn\\
&&\sum_v (h_v-h) (\sum_i |P_{v_i}|-|P_v|)=\sum_v (h_v-h_{v'})(4 m^4_v-|P_v|+2 n^A_v)
\eea
where $m^4_v$ is the number of end-points following $v$, $n^A_v$ is the number of external $A$ lines we get
\bea
&&\int dx_{v_0} |W_{\t,\bP,T}(x_{v_0})|\le L^4 
\g^{-h[-4+{3|P_{v_0}|\over 2}-2 n+n^J_{v_0}]}\nn\\
&&\prod_{v\,\hbox{\ottorm not e.p.}} \left\{ {1\over s_v!}
C^{\sum_{i=1}^{s_v}|P_{v_i}|-|P_v|}
\g^{-(-4+{3|P_v|\over 2}-2 m_{4,v}+n^A_v+z_v)}\right\}[\g^{-2 N n}] [\g^N/M]^{6 n}
\nn\eea
and finally
\be
\int d x_{v_0} |W_{\t,\bP,T}(\xx_{v_0})|\le L^4 
\g^{-h d_{v_0}} C^n|\l|^{n}
[\prod_{\tilde v} \
{1\over s_{\tilde v}!}\g^{-d_{\tilde v} (h_{\tilde v}-h_{\tilde v'}) }
][\prod_{\tilde v} \g^{-2(N-h_{\tilde v})\bar m_{\tilde v})}]
 [\g^N/M]^{6 n} 
\ee
where: $\tilde v\in \tilde V$ are the vertices on the tree such that $\sum_i |P_{v_i}|-|P_v|\not=0$, $\tilde v'$ is the
vertex in $\tilde V$ immediately preceding $\tilde v$ or the root; $\bar m_v$ is the number of normal end-point following $\tilde v$ and not any following vertex $\tilde v\in \tilde V$;
$d_v=-4+{3|P_v|\over 2}+n^A_v+z_v$.
Finally the number of addenda in $\sum_{T\in {\bf T}}$ is bounded by
$\prod_{v} s_v!\;
C^{\sum_{i=1}^{s_v}|P_{v_i}|-|P_v|}$. 
In order to bound the sums over the scale labels and $\bP$ we first use
the inequality
\be
\prod_{\tilde v} \g^{-d_{\tilde v} (h_{\tilde v}-h_{\tilde v'}) }
\le [\prod_{\tilde v} \g^{-{1\over 2}(h_{\tilde v}-h_{\tilde v'})}]
[\prod_{\tilde v}\g^{-{3|P_{\tilde v}|\over 4}}]
\ee
where $\tilde v$ are the non trivial vertices, and $\tilde v'$ is the
non trivial vertex immediately preceding $\tilde v$ or the root. The
factors $\g^{-{1\over 2}(h_{\tilde v}-h_{\tilde v'})}$ in the r.h.s.
allow to bound the sums over the scale labels by $C^n$.
Finally if there if there is at least a normal end-point the bound improves by a factor 
$\g^{\th (h-N)}$ as $\bar m_{\tilde v} \ge 1$ for some $\tilde v$ so  that, if 
\be
\prod_{\tilde v} \g^{-d_{\tilde v} (h_{\tilde v}-h_{\tilde v'}) }
 [\prod_{\tilde v} \g^{-2(N-h_{\tilde v})\bar m_{\tilde v})}]\le \g^{\th (h-N)}
\prod_{\tilde v} \g^{-\hat d_{\tilde v} (h_{\tilde v}-h_{\tilde v'}) }
\ee
with $\hat d_v=d_v-\th>0$ so that the sum over scales can be still done. This  completes the proof of \pref{bb}.

\section{The three current correlation}

We have now to prove Theorem 2.
In order to compute $S^3(x,x_1,x_2)=<\tilde j_{B, \m,x};\tilde j_{A^{\e_1},\m_1,x_1};
\tilde j_{A^{\e_2},\m_2,x_2}>$ we perform the derivatives of $\WW(A)$ given by \pref{sac} 
with respect to $B_{\m,x}, A^{\e_1}_{\m_1,x_1},A^{\e_2}_{\m_2,x_2}$. The result can be written as 
$S^3=S^3_a+S^3_b$, where $S^3_a$ is given by trees with only special end-points, and $S^3_b=\sum_h \sum^*_\t 
S_\t(x,x_1,x_2)$ where $\sum^*_\t$ is sum over trees with at least one normal end-point, see Fig. 5.
The renormalized triangle graphs have the form, in the case of three $B$ currents
\be \sum_{h_1,h_2,h_3}[\prod_{j=1}^3 {Z^J_{h_j,s,i}\over Z_{h_i,s,i} }]
\int {dk \over (2\pi)^4}
{\rm Tr} {f^{h_1}(k)\over -i  \s^s_\m k_\m}(-i\s^s_\m) {f^{h_2}(k+p)\over -i \s^s_\m (k_\m+p_\m)}(-i \s^s_\n){f^{h_3}(k+p^2)\over -i   \s^s_\m (k_\m+p^2_\m)}(-i \s^s_\r)+[(\n, p_1)\to (\s,p_2)]
\ee
and a similar expression holds for the $BWW$ currents.
The main difference with respect to the triangle graphs seen in the non-interacting case
is that the wave and the vertex are non trivial function of the momentum scale. We can use now \pref{sep1} to further decompose the triangle graph as a sum of two terms, one
in which $Z_{s,i,h}$, $Z^J_{s,i,h}$ ,  $Z^W_{a,h}$ are replaced by $1$ and an extra term. The sum of such extra term
plus  $S^3_b$
constitute the term $R^{1,B}_{\m,\n,\r}, R^{1,W}_{\m,\n,\r}$ in \pref{sec1}. The other term is exactly coinciding with the non-interacting case using that
$\sum_{h=-\io}^N f^h=\chi_N$.

\insertplot{530}{79}
{\ins{50pt}{40pt}{$=$}
\ins{130pt}{40pt}{$+$}
}
{figjsp44a}
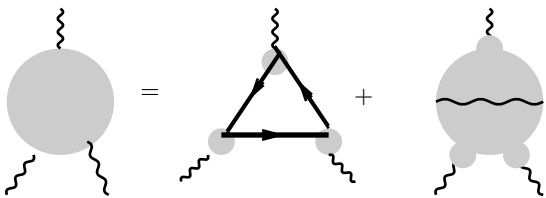
{\label{h2} Graphical representation to $S_3$. The first term in the r.h.s. represents $S^3_a$ and is a sum of triangle graphs; the dots 
represent the renormalization $Z^W_{h}$ or $Z^J_{h}$. The other term represents $S^3_b$, which is sum of terms with at least a $\l$ vertex.
} {0}
Let us consider $S^3_b$.
With respect to the bound obtained in the previous section, we have to take into account that there is no contribution from the integrals over the coordinates.
Given a tree $\t$, we can associate a tree $\t^*$, which is the tree obtained by $\t$ 
by erasing all the vertices not necessary to connect the special end-points; given a non-trivial vertex $v\in\t^*$, we call $x^*_v$ the coordinates
associated to end-points in $\t^*$ following $v$, and $\d_v$ is the length of the shortest tree graph connecting the points $x^*_v$. The number of non trivial vertices $v\in \t^*$ is $\le 3$.
The lack of integration over the external coordinates gives an extra factor $\prod_v \g^{4 h_v(S^*_v-1)h_v }$  where $S^*_v$ is the number of branches in $\t^*$ following $v$
(each integration contribute with a 
factor $\g^{-4 h_v(S^*_v-1)h_v }$); moreover we can write 
\be
e^{-\sqrt{\g^h |x|}}\le e^{-{1\over 2}\sqrt{\g^h |x|}} \prod_{k=-\io}^0 e^{-c\sqrt{\g^h \g^{k} |x|}}= e^{-{1\over 2}\sqrt{\g^h |x|}} \prod_{k=-\io}^h e^{-c\sqrt{\g^{k} |x|}}
\ee
The first factor is used to perform the integrations and from the second we get a factor $e^{-(\g^{h_v} |\d_v|)^{1\over 2}} ]$ for any non trivial vertex in $v$
so that, if $m=3$ and $n$ is the number of normal end-points ($\prod_{\tilde v} \g^{-\hat d_{\tilde v} (h_{\tilde v}-h_{\tilde v'}) }=\prod_v \g^{-\bar d_v}$)
\be
|S_\t(x,x_1,x_2)|\le C^n |\l|^n 
 \g^{-h (-4+2 m)}\g^{\th(h_{v_0^*}-N )}\\
[\prod_v \g^{-\hat d_v}] 
[ \prod_{n.t. v \in \t^*} \g^{4 h_v(S^*_v-1)h_v } e^{-(2^{h_v} |\d_v|)^{1\over 2}} ]
\label{bo2} 
\ee
with $v_0^*$ is the first non trivial vertex in $\t^*$.
%
%
%
The factor $\g^{-h (-4+2 m)}$ apparently forbids to sum over $h$, and we need to use the decay factors associated to the propagators. In order to do that
we can write \be \g^{-h (-4+2 m)}\prod_{v_0\le v\le v_0^*}\g^{-\hat d_v}
=\g^{-h_{v_0^*} (-4+2 m)}\prod_{v_0\le v\le v_0^*}\g^{-\tilde d_v}\ee
where $m_v=n^A_v$ for $v_0\le v\le v_0^*$ and $\hat d_v-(-4+n^A_v)=
\tilde d_v=3/2|P_v|+z_v-\th>0$.
We call $S^1_{v_0^*}$ the branches connecting to special $A$ end-points. Using that $S_{v_0^*}=S^1_{v_0^*}+S^2_{v_0^*}$
and 
$
n^A_{v_0^*}=S^1_{v_0^*}+\sum_{i=1}^{S^2_{v_0^*}} n^A_{v_i}$, $n^A_{v_i}$ the number of special end-points in $\t_i^*$ we have
\be
-(-4+2 n^A_{v_0^*})+4(S_{v_0^*}-1)=-2 S^1_{v_0^*}-\sum_{i=1}^{S^2_{v_0^*}} 2 n^A_{v_i}+4 S^1_{v_0^*}
+4 S^2_{v_0^*}
=2 S^1_{v_0^*}-\sum_{i=1}^{S^2_{v_0^*}}(-4+2 n^A_{v_i})\ee
Therefore to the vertex $v_0^*$ is associated 
$
\g^{\th h_{v_0^*} } \g^{2 S^1_{v_0^*} h_{v_0^*} }
$.
We can repeat the same argument on each of the subtrees $\t^*_1,..,\t^*_{S^2_{v_0^*}}$;
%
%
moreover $\g^{2 S^1_v h_v} 
 e^{-(2^{h_v} |\d_v|)^{1\over 2}}\le C_\d$ as $S^1_v\le 3$ so that 
\be
|S_\t(x,x_2,x_3)|\le C^n |\l|^n 
\g^{-\th N } [\g^{\th h_{v_0^*} } \g^{2 S^1_{v_0^*} h_{v_0^*} }
e^{-(2^{h_{v_0^*}} |\d|)^{1\over 2}} ] 
[\prod_v \g^{-\bar d_v}] 
\label{bo3} 
\ee
where $\bar d_v=\tilde d_v$ if $v\in \t^*$, $\bar d_v=\hat d_v$ otherwise.
The sum over the scale difference is done using that $\bar d_v>0$; the remaining sum is done using that
if $\d\equiv \g^{-h_\d}$, if $1\le S^1\le 3$
\be
\sum_h \g^{(\th+2 S^1) h } e^{-(2^{h} |\d|)^{1\over 2}}=\g^{(\th+S^1) h_\d}\sum_h \g^{(\th+2 S^1) (h-h_\d) } e^{-\g^{(h-h_\d)/2}}
\le  C_\d\ee
uniformly in $N$. In conclusion a bound $|\l|({1\over \g^N })^\th C_\d$ is found. A similar bound is found for the corrections coming from the first term, as they have an extra factor $\g^{\th(h-N) }$ 
from \pref{sep1}. This concludes the proof of the bound \pref{sec}.
\vskip.5cm

{\bf Acknowledgements.} This work has been supported 
by MIUR, PRIN 2017 project MaQuMA, PRIN201719VMAST01.


\begin{thebibliography}{999999}

\bibitem{R}
J. A. Minahan, P. Ramond, and R. C. Warner
{\it 
Phys. Rev. D }
41, 715  (1990)
\bibitem{R1}
L. Alvarez-Gauma'  An Introduction to Anomalies. In: Velo G., Wightman A.S. (eds) Fundamental Problems of Gauge Field Theory. NATO ASI Series (Series B: Physics), vol 141. Springer, Boston (1986)


\bibitem{T}
G. ’t Hooft, The Background Field Method in Gauge Field Theories, in: Functional and Probabilistic Methods in Quantum Field Theory.
Proceedings, 12th Winter School of Theoretical Physics, Karpacz, 345–369 (1975)


\bibitem{AB} S. L. Adler, W. A. Bardeen. {\it Phys. Rev.} 182, 1517 (1969).

\bibitem{F}
K. Fujikawa,  {\it Phys. Rev. Lett. }
42, 1195-1198 (1979)

\bibitem{K} C. Kopper, B. Leveque.
{\it 
Jour. of Math. Phys.}   53, 022305 (2012)

\bibitem{AB1} S. Adler Fifty Years of Yang-Mills Theory, G. 't Hooft editor World Scientific (2005)

\bibitem{A0} M. Luscher, {\it  Nucl. Phys. B} 549, 295 (1999) [arXiv:hep-lat/9811032].

\bibitem{A01}
M. Luscher, {\it  Nucl. Phys. B} 568, 162 (2000)

\bibitem{A1} D. Kadoh, Y. Kikukawa   {\it JHEP} 0805:095 (2008)

\bibitem{A2}
 D. M. Grabowska, D. B. Kaplan  {\it Phys. Rev. Lett..} 116, 211602 (2016)

\bibitem{A3} M. Golterman
 {\it Nucl.Phys.Proc.Suppl.} 94 , 189-203 (2001)

\bibitem{Pr}
J.Preskill
{\it Annals Phys.} 210, 323-379 (1991)

\bibitem{Pr1}
J. Soto
{\it  Phys. Rev. D} 45, 4621 (1992)


\bibitem{P0}  S. Coleman, S. L. Glashow. {\it 
Phys.Rev. D}  59, 116008,1999

\bibitem{Pol} 
A. M. Polyakov. Gauge fields and strings.
{\it  Contemp.Concepts Phys} 3, 1-301 (1987)

\bibitem{Po}
J. Polchinski. Recent directions in particle theory  {\it Boulder proceedings } 235-274  (1992)

\bibitem{Wi} E. Witten. {\it Nature Physics} 14, 116–119 (2018)

\bibitem{M1} V. Mastropietro Non Perturbative Renormalization 1-304 World Scientific (2008)

\bibitem{M11} V. Mastropietro . {\it J.Math.Phys.} 48, 022302 (2007)

\bibitem{BFM1} G.Benfatto, P.Falco, V.Mastropietro {\it
Comm.Math. Phys.}  273,  1, 67--118 (2007);  {\it Comm. Math. Phys.}  285, 2,
713--762 (2009)
\bibitem{BFM2} G.Benfatto, P.Falco, V.Mastropietro
{\it Phys. Rev. Lett. } 104, 075701 (2010)



\bibitem{GMP} A. Giuliani, V.Mastropietro, M.Porta arXiv:1907.00682 

\bibitem{GMP1} V. Mastropietro
{\it JHEP} 2020
  
\bibitem{NN} H. B. Nielsen, M. Ninomiya. {\it Phys. Lett. B}, 130, 389 (1983).












\end{thebibliography}
\end{document}